\documentclass[12pt]{emulateapj}
\usepackage{epsfig}

\newcommand{\ngc}{NGC~4258}
\newcommand{\sgra}{Sgr A$^*$}

\newcommand{\beppo}{\textsl{BeppoSAX}}

\def\spose#1{\hbox to 0pt{#1\hss}}

\def\multleft#1{\hbox to size{\vbox {\halign {\lft{##}\cr #1}}\hfill}\par}
\def\multright#1{\hbox to size{\vbox {\halign {\rt{##}\cr #1}}\hfill}\par}

\def\boxit#1{\vbox{\hrule\hbox{\vrule\kern3pt\vbox{\kern3pt
          #1 \kern3pt}\kern3pt\vrule}\hrule}}

\def\cm{{\rm\thinspace cm}}

\def\erg{{\rm\thinspace erg}}
\def\eV{{\rm\thinspace eV}}

\def\keV{{\rm\thinspace keV}}
\def\km{{\rm\thinspace km}}

\def\Msun{\hbox{$\rm\thinspace M_{\odot}$}}
\def\pc{{\rm\thinspace pc}}

\def\s{{\rm\thinspace s}}

\def\ergpcmsqps{\hbox{$\erg\cm^{-2}\s^{-1}\,$}}

\def\ergps{\hbox{$\erg\s^{-1}\,$}}

\def\kmps{\hbox{$\km\s^{-1}\,$}}

\def\pcmsq{\hbox{$\cm^{-2}\,$}}

\newcommand\aproxgt{\mathrel{%
      \rlap{\raise 0.511ex \hbox{$>$}}{\lower 0.511ex \hbox{$\sim$}}}}
\newcommand\aproxlt{\mathrel{%
      \rlap{\raise 0.511ex \hbox{$<$}}{\lower 0.511ex \hbox{$\sim$}}}}



\begin{document}

\title{Probing the accretion disk and central engine structure of
NGC~4258 with {\it Suzaku} and {\it XMM-Newton} observations}

\author{Christopher~S.~Reynolds\altaffilmark{1}, Michael
A. Nowak\altaffilmark{2}, Sera Markoff\altaffilmark{3}, Jack
Tueller\altaffilmark{4}, Joern Wilms\altaffilmark{5}, Andrew
J. Young\altaffilmark{6}}

\altaffiltext{1}{Department of Astronomy and the Maryland Astronomy Center for Theory and Computation, University of Maryland, College Park, MD 20742-2421} 
\altaffiltext{2}{Kavli Institute for Astrophysics and Space Research, Massachusetts Institute of Technology, 77 Massachusetts Avenue, Cambridge, MA 02139} 
\altaffiltext{3}{Astronomical Institute ``Anton Pannekoek,'', Kruislaan 403, 
NL-1098 SJ Amsterdam, Netherlands}
\altaffiltext{4}{NASA/Goddard Space Flight Center, Astrophysics Science Division, Greenbelt, MD 20771}
\altaffiltext{5}{Dr. Karl Remeis-Sternwarte, Astronomisches Institut, University of Erlangen-Nuremberg, Sternwartstrasse, 96049 Bamberg, Germany}
\altaffiltext{6}{Department of Physics, University of Bristol, Tyndall Avenue, Bristol BS8~1TL, UK.}

\begin{abstract}
We present an X-ray study of the low-luminosity active galactic
nucleus (AGN) in NGC~4258 using data from {\it Suzaku}, {\it
XMM-Newton}, and the {\it Swift}/BAT survey.  We find that signatures
of X-ray reprocessing by cold gas are very weak in the spectrum of
this Seyfert-2 galaxy; a weak, narrow fluorescent-K$\alpha$ emission
line of cold iron is robustly detected in both the {\it Suzaku} and
{\it XMM-Newton} spectra but at a level much below that of most other
Seyfert-2 galaxies.  We conclude that the circumnuclear environment of
this AGN is very ``clean'' and lacks the Compton-thick obscuring torus
of unified Seyfert schemes.  From the narrowness of the iron line,
together with evidence for line flux variability between the {\it
Suzaku} and {\it XMM-Newton} observations, we constrain the line
emitting region to be between $3\times 10^3r_g$ and $4\times 10^4r_g$
from the black hole.  We show that the observed properties of the iron
line can be explained if the line originates from the surface layers
of a warped accretion disk.  In particular, we present explicit
calculations of the expected iron line from a disk warped by
Lens-Thirring precession from a misaligned central black hole.
Finally, the {\it Suzaku} data reveal clear evidence for large
amplitude 2--10\,keV variability on timescales of 50\,ksec as well as
smaller amplitude flares on timescales as short as 5--10\,ksec.  If
associated with accretion disk processes, such rapid variability
requires an origin in the innermost regions of the disk ($r\approx
10r_g$ or less).  Analysis of the difference spectrum between a high-
and low-flux state suggests that the variable component of the X-ray
emission is steeper and more absorbed than the average AGN emission,
suggesting that the primary X-ray source and absorbing screen have
spatial structure on comparable scales.  We note the remarkable
similarity between the circumnuclear environment of NGC~4258 and
another well-studied low-luminosity AGN, M81*.
\end{abstract}


\section{Introduction}

The low-luminosity active galactic nucleus (LLAGN) \ngc\ is an
essential object in our quest to understand the astrophysics of
extragalactic supermassive black holes.  Very Large Array (VLA) and
Very Long Baseline Interferometry (VLBI) observations have found a set
of water masers that trace a nearly edge-on, geometrically-thin gas
disk $\sim 0.2-0.3\pc$ from the central black hole (Miyoshi et
al. 1995; Herrnstein et al. 1999).  The near-perfect Keplerian
velocity curve of these water masers provides one of the strongest and
most robust pieces of evidence for the existence of extragalactic
supermassive black holes (Miyoshi et al. 1995).  Furthermore, the
dynamics of this disk allows precise measurements of the central black
hole mass, (outer) accretion disk inclination and warping, \emph{and}
its distance ($M = 3.9 \pm 0.3 \times 10^{7}\,\Msun$, $D = 7.2 \pm
0.5$\,Mpc, yielding $r_g \equiv GM/c^2 = 5.8\times 10^{12}\,{\rm cm}$,
1\,pc $=5.4 \times 10^5~r_g$, and 1'' = 35\,pc; Miyoshi et al. 1995;
Herrnstein et al. 1999). This is the only AGN for which the black hole
mass, distance, and (outer) accretion disk geometry are so accurately
known.

Due to the precision with which the black hole mass and distance are
known, \ngc\ can be used as a basic test-bed for our models of black
hole accretion.  The overall luminosity of the AGN is small compared
with the Eddington luminosity of the black hole, $L\sim
10^{-4}\,L_{\rm Edd}$, the cause of which has remained
controversial. Is the small luminosity simply due to a very small mass
accretion rate through a radiatively-efficient disk, as suggested by
modeling the physics of the maser production Neufeld \& Maloney
(1995)?  Or does the disk make a transition to a
radiatively-inefficient accretion flow (RIAF) at some radius as
suggested by the modeling of Lasota et al. (1996) and Gammie, Narayan
\& Blandford (1999)?  What is the role of the jet, and how much of the
radiative luminosity of the AGN is actually due to the jet (Yuan et
al. 2002).  In the bigger picture, what is the fundamental difference
among the accretion flows in the most underluminous galactic nuclei
(e.g., \sgra), LLAGN and powerful AGN?

Sensitive X-ray observations provide a powerful means of probing both
large scale and small scale structures within NGC~4258 and hence
addressing these questions.  Soft X-ray thermal emission from hot gas
associated with the well known helically twisted jets (the anomalous
arms) has been known since the {\it Einstein} days (Fabbiano et
al. 1992; also see {\it ROSAT} work of Pietsch et al.  1994; Cecil,
Wilson \& De Pree 1995; Vogler \& Pietsch 1999).  However, power-law
X-ray emission from the (absorbed) central engine of the AGN itself
was not seen until the advent of {\it ASCA} (Makishima et al. 1994;
Reynolds, Nowak \& Maloney 2000).  {\it ASCA} clearly revealed
variability of both the absorbing column and hard X-ray flux on the
timescale of years (Reynolds et al. 2000; hereafter R00), a result
supported by short observations of NGC~4258 by {\it XMM-Newton} and
{\it Chandra} (Pietsch \& Read 2002; Young et al. 2004; Fruscione et
al. 2005).  The most sensitive hard X-ray ($>10\keV$) study of
NGC~4258 was conducted by \beppo\ (200\,ksec; Fiore et al. 2001) which
detected the AGN emission out to beyond 50\,keV (Fiore et al. 2001).
\beppo\ also revealed day-timescale variability of the hard X-ray
power-law, setting a firm upper limit of $250\,r_g$ ($5 \times
10^{-4}$\,pc) to the X-ray emission region size.

In this {\it Paper}, we present results from new {\it Suzaku} and {\it
XMM-Newton} observations of NGC~4258 which, when supplemented with
survey data from the {\it Swift} Burst Alert Telescope (BAT), gives us
an unprecedented view of this AGN from 0.3\,keV upto 140\,keV.  These
data suggest a circumnuclear environment that is remarkably ``clean''
compared with other Seyfert 2 nuclei.  {\it Suzaku} also reveals rapid
variability of the AGN emission, allowing us to set new constraints on
the size/compactness of the X-ray source.  The plan of this paper is
as follows.  Section~2 briefly discusses the data that we utilize and
the basic reduction steps.  We present our analysis of the spectrum,
as well as spectral variability, in Section~3.  Section~4 discusses
the implications of these results for our understanding of the
structure of this AGN and the origin of the X-ray emission.  In
particular, we argue that the iron line has all of the properties
expected if it were to originate from the surface layers of the
(warped) accretion disk.  Throughout this paper we quote error bars at
the 90\% confidence level for one interesting parameter.  All error
bars on figures are displayed at the $1\sigma$ level.

\section{Observations and data reduction}

In this section, we discuss the new observations presented in this
paper and the subsequent data reduction.   

{\it Suzaku} observed NGC~4258 for a total of 186\,ksec starting
10-Jun-2006 as part of the Cycle-1 Guest Observer Program (PI: Itoh,
US Co-PI: Reynolds).  All four X-ray Imaging Spectrometers (XIS~0--3)
as well as the Hard X-ray Detector (HXD) were operational and
collecting data, and NGC~4258 was placed at the ``nominal HXD''
aimpoint.  Reduction started from the cleaned Version-2 data products,
and data were further reduced using FTOOLS version 6.4 according to
the standard procedure outlined in the ``Suzaku Data Reduction (ABC)
Guide''.  The standard filtering resulted in 88.5\,ksec of ``good''
XIS data.  Spectra and lightcurves were extracted from all XISs using
a circular region of radius 3.25\,arcmin centered on NGC~4258.
Background spectra were obtained from rectangular source free regions
(avoiding the calibration sources) around the chip edges.  Response
matrices and effective area curves were generated using the {\tt
xisrmfgen} and {\tt xissimarfgen} tools, respectively, using the
recommended 400,000 photons per energy bin during the construction of
the effective area files.  We also utilize HXD data in this paper.
Standard filtering resulted in 97.5\,ksec of ``good'' {\it
Suzaku}-HXD/PIN data from which a spectrum was constructed.  A PIN
background spectrum was produced that included the Cosmic X-ray
Background (CXB) plus the latest model of the detector background.  We
do not consider HXD/GSO data in this paper due to the fact that the
high background of this detector makes a detection of NGC~4258
impossible.

The {\it XMM-Newton} data presented in this paper result from a
continuous 65\,ksec exposure started on 17-Nov-2006, 160 days after
the {\it Suzaku} observation.  All of the European Photon Imaging
Cameras (EPIC) were operated in {\tt PrimeFullWindow} mode, and the
data were cleaned using the Science Analysis System (SAS) version 7.1
following the standard procedure outlined in the ``User's Guide to
XMM-Newton SAS''.  We reject data during three background flares (PN
count rate $>$50\,ct\,s$^{-1}$).  The final ``good'' exposure time for
the EPIC detectors is 55\,ksec.  EPIC spectra were extracted using a
circular extraction region of radius 30\,arcsec centered on the bright
nucleus of NGC~4258, and background spectra were extracted using a
nearby source free circular region of radius 1.5\,arcmin.

The {\it Swift/BAT} is a wide field (2 steradians) coded aperture hard
X-ray instrument which, during normal operations, surveys 60\% of the
sky each day at $<$20\,milliCrab sensitivity.  The BAT spectrum used
here was prepared as part of the 22-month extension to the BAT-survey.
The BAT survey spectra are derived from an independent all sky mosaic
map in each of nine energy bins averaged over 22 months of data
beginning on Dec 5 2004.  The energy bin edges are 14, 20, 24, 35, 50,
75, 100, 150, 195 keV.  The nature of the coded-mask analysis
naturally results in a background-subtracted source spectrum.  As
discussed in Tueller et al. (2008), fitting of the BAT data was
performed using a diagonal response matrix which correctly accounts
for instrumental systematics in sources with spectral indices similar
to the Crab (photon index $\Gamma\sim 2$).  See Tueller et al. (2008)
for more details.

All spectra were binned to a minimum of 20 counts per bin to
facilitate $\chi^2$ fitting.  All spectral analysis presented here is
performed using XSPECv11.3.2.

\section{Results}

In this section, we describe the spectral and temporal properties of
the X-ray emission from NGC~4258.  Given its superior signal-to-noise
and energy band our discussion focuses on the {\it Suzaku} dataset,
although we compare and contrast with the new {\it XMM-Newton} dataset
where ever appropriate.  In order to extend the energy reach of our
time-averaged spectral study (\S\ref{sec:broadspec}), we supplement
the {\it Suzaku} data with {\it Swift}/BAT data.  Finally, in order to
study long-term variations in the all-important fluorescent iron line,
we include data from our new {\it XMM-Newton} observation as well as
archival {\it XMM-Newton} and {\it ASCA} data.

\subsection{The 0.3--140\,keV spectrum}
\label{sec:broadspec}

\begin{figure}
\centerline{
\psfig{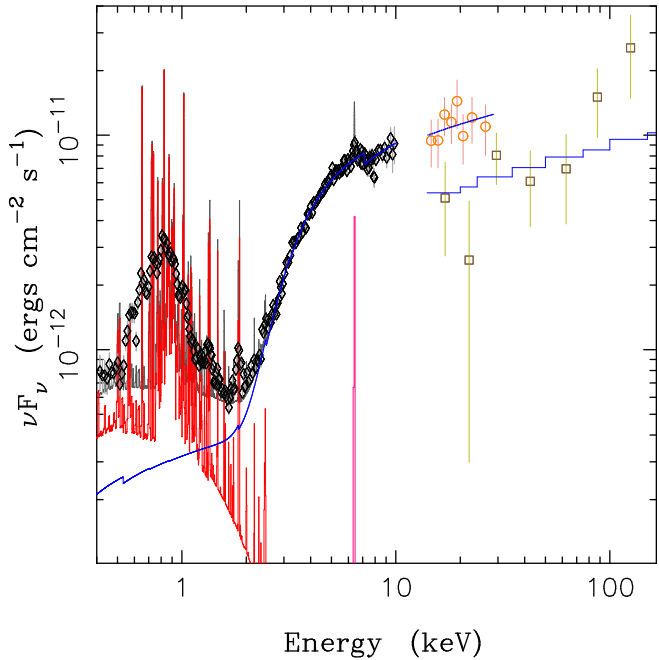}
}
\caption{The 0.3--140\,keV spectrum of NGC~4258.  Shown here are data
from {\it Suzaku}/XIS1 (0.3--10\,keV; black), {\it Suzaku}/PIN
(14--30\,keV; orange), and the {\it Swift}/BAT survey (15--140\,keV;
gold).  The spectral model consists of two optically-thin thermal
plasma component (magenta; unabsorbed apart from the effects of the
cold Galactic absorption column of $N_{\rm Gal}=1.45\times
10^{20}\,{\rm cm}^{-2}$), an absorbed power-law (with intrinsic
absorption $N_H=9.2\times 10^{22}\,{\rm cm}^{-2}$ and photon index
$\Gamma=1.77$), and a narrow 6.4\,keV iron fluorescence line.}
\label{fig:spectrum}
\vspace{0.3cm}
\end{figure}

The combination of the {\it Suzaku}/XISs, the {\it Suzaku}/PIN and the
{\it Swift}/BAT allows us to form the spectrum of \ngc\ over the
energy range 0.3--140\,keV (see Fig.~\ref{fig:spectrum} which, for
clarity, only shows one of the four XISs).  This spectrum extends to
significantly higher energy than any previous study of \ngc.  It must
be cautioned, however, that the {\it Swift}/BAT data are collected
over a 22-month period compared with the {\it Suzaku} ``snapshot''.
In this study, we make the assumption that the form/shape of the
high-energy ($>10\keV$) spectrum does not change with time even if its
normalization does change.  This allows us to perform joint spectral
fits of the {\it Suzaku} and {\it Swift}/BAT data in order to study
the nature of the X-ray source and the circumnuclear environment.

Guided by previous studies of this source, we model this spectrum as
the superposition of optically-thin thermal plasma emission with
temperature $T$ (described by the XSPEC model {\tt vmekal}; Mewe et
al. 1985; Kaastra 1992; Liedahl et al. 1995), an absorbed power-law
(with photon index $\Gamma$ absorbed by a cold column $N_{\rm H}$),
and an additional continuum component required to correctly describe
the inflection point in the spectrum around 2\,keV where thermal
plasma emission and absorbed power-law swap dominance.  There are
several possible identifications of this additional continuum
component including (1) AGN emission that has scattered around the
absorbing matter, (2) AGN emission that has leaked through a patchy
absorber, (3) hard X-ray emission associated with X-ray binaries in
the galaxy, or (4) thermal emission from very hot gas associated with
star formation or the interaction of the AGN jet with the galactic
disk.  Both the intrinsic absorption and the Galactic absorption (from
a column $N_{\rm H}=1.45\times 10^{20}\,{\rm cm}^{-2}$; Dickey \&
Lockman 1990) were described using the {\tt phabs} model.  To obtain a
good description of the soft X-ray data, we require that the elemental
abundances are allowed to vary relative to solar values (as defined by
Anders \& Grevesse 1989) in two groups, Group A (with abundance $Z_A$)
consisting of \{C,N,O,Ne,Na,Mg,Al,Si,S,Ar,Ca\}, and group B (with
abundance $Z_B$) containing \{Fe,Ni\}.

In our Canonical Spectral Model, the additional continuum component is
modeled by a power-law component with photon index equal to that of
the main (absorbed) AGN powerlaw ($\Gamma_2=\Gamma$) that is
unaffected by the intrinsic absorption.  This is an appropriate
spectral model to describe the scattering or leaky absorber scenario.
This model provides a decent description of the spectrum ($\chi^2/{\rm
dof}=4506/4003$) with best-fitting parameters
$\Gamma=1.75^{+0.05}_{-0.04}$, $N_{\rm H}=(9.2^{+0.4}_{-0.3})\times
10^{22}\,{\rm cm}^{-2}$, $kT=0.54\pm 0.01\keV$,
$Z_A=0.49^{+0.10}_{-0.08}Z_\odot$, $Z_B=0.27^{+0.05}_{-0.04}Z_\odot$.
The normalization of the additional continuum component relative to
the main AGN powerlaw is $f=6.0\pm 0.4\%$; this can be interpreted as
the scattering or leakage fraction.  If we allow the photon index of
the additional continuum component to be free, the fit does not
improve and we find that $\Gamma_2$ is very poorly constrained.  Using
this spectral model, we deduce an observed 0.5--10\,keV flux $F^{\rm
obs}_{0.5-10}=9.7\times 10^{-12}\ergpcmsqps$ and observed 0.5--10\,keV
luminosity $L^{\rm obs}_{0.5-10}=6.0\times 10^{40}\ergps$.  Removing
the effects of absorption within the model implies an intrinsic
0.5--10\,keV luminosity of $L^X_{0.5-10}=1.4\times 10^{41}\ergps$.
The best-fitting normalization of the BAT model is 48\% that of the
{\it Suzaku} model.  We interpret this as true variability, i.e., the
{\it Suzaku} observation caught the source at a time when the
high-energy emission has twice the normalization of the 22-month
average spectrum.

The second continuum component can also be described by an additional
thermal plasma component; however, while a good fit can be obtained
($\chi^2/{\rm dof}=4501/4004$), a rather hot ($kT>5.5\keV$) and
low-metallicity ($Z<0.26Z_\odot$) plasma is required.  Thus, since it
is not clear from where such a plasma component would originate, we
prefer the power-law description of this additional continuum
component.

Our Canonical Spectral Model fit leaves a line-like residue in the
0.5--0.6\,keV range that likely signals soft X-ray complexity beyond
the simple one-temperature thermal plasma model.  Adding a second
plasma component with an identical abundance pattern but different
temperature ($kT=0.22\pm 0.02\keV$) leads to a significant improvement
in the goodness of fit ($\chi^2/{\rm dof}=4360/4001$).  However, a
robust exploration of these multi-temperature solutions is not
possible at XIS-resolutions (e.g., the abundances are poorly
constrained) and we defer further discussion of this to a future
publications in which high resolution spectral data from the {\it
  XMM-Newton}/RGS and a long {\it Chandra}/HETG observation) will be
discussed.

The combination of the XIS, PIN and BAT data allows us to examine the
hard X-ray spectrum of this source in more detail than previously
possible.  For the hard-band study of this paragraph, the data below
3\,keV were formally excluded from the analysis to prevent their
high-statistical weight from biasing the high-energy fit.  No
significant deviations from a pure absorbed power-law are detected
above 3\,keV.  Of course, in order for the observed powerlaw (with
$\Gamma<2$) not to possess a divergent energy, it must cut-off or roll
over at some high energy.  If the power-law has an exponential cutoff,
the characteristic e-folding energy is constrained by our high-energy
data to be $E_{\rm fold}>124\keV$.  If we instead assume a pure-power
law with cold X-ray reflection (modeled using the {\tt pexrav} code;
Magdziarz \& Zdziarski 1995), the constraints on the ``reflection
fraction'' ${\cal R}$ are very dependent upon the assumed inclination
of the reflector.  If the reflector has a slab geometry with a very
high inclination ($i>80^\circ$), as expected if we identify it with
the inner accretion disk, then these data provide no meaningful
constraints on the reflection fraction.  For more face-on reflection
(e.g., the surfaces of discrete cold clouds or the inner wall of a
cold torus on the far side of the X-ray source), we find ${\cal
R}<0.43$.  If we allow both reflection and an exponential cutoff
simultaneously, the limits on the e-folding energy become weaker
($E_{\rm fold}>67\keV$) but constraints on reflection are essentially
unaffected.  The implications of this result for the circumnuclear
environment in this Seyfert nucleus is discussed in \S4.

An identical analysis of the 0.7--10\,keV {\it XMM-Newton}/EPIC (PN
and MOS) data (supplemented with the {\it Swift}/BAT spectrum) gives a
similar picture, although there are some quantitative differences.
Firstly, the flux of the thermal plasma emission is lower by a factor
of two, a simple consequence of the fact that much of this emission
lies in the anomalous arms and is outside of the our EPIC extraction
region.  The temperature and iron abundance of this emission
($kT=0.58\pm 0.01\keV$, $Z_B=0.17^{+0.08}_{-0.05}Z_\odot$) are very
similar to that derived from {\it Suzaku}, although the light metal
abundance is slightly lower ($Z_A=0.20^{+0.11}_{-0.06}Z_\odot$).  The
most robust differences are with the parameters describing the
absorbed power-law; the power-law is flatter ($\Gamma=1.65\pm 0.07$)
and less absorbed [$N_H=(7.7\pm 0.5)\times 10^{22}\pcmsq$] in the new
{\it XMM-Newton} data as compared with the earlier {\it Suzaku}
observation.  Accompanying these changes is a decrease in intrinsic
(unabsorbed) 0.5--10\,keV luminosity from $L^X_{0.5-10}=1.4\times
10^{41}\ergps$ to $L^X_{0.5-10}=6\times 10^{40}\ergps$.  We will see
below that these long-term changes are in the same sense as short term
variability seen within the {\it Suzaku} observation and may be
revealing aspects of the spatial structure of the X-ray source and
absorber.

\subsection{The Suzaku detection of a weak iron line}

\begin{figure}
\centerline{
\psfig{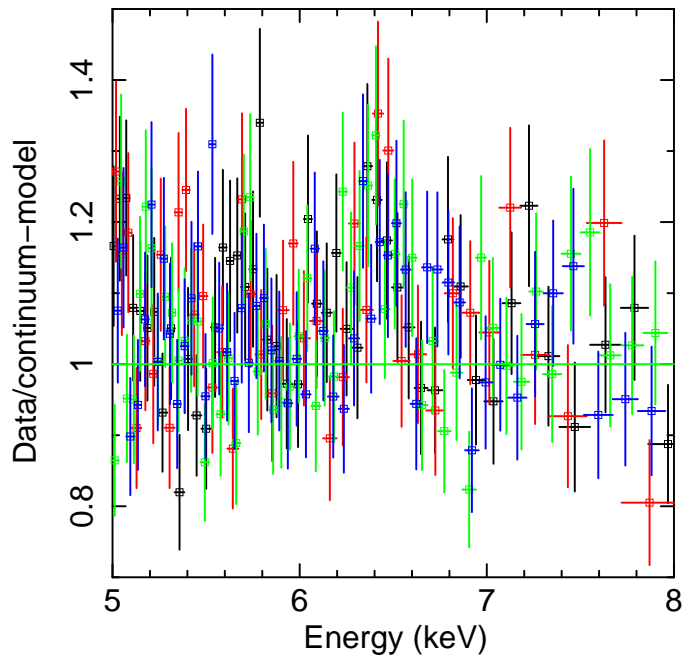}
}
\caption{XIS residues from the best-fitting continuum model showing
the presence of a 6.4\,keV fluorescent line of cold iron.  Data from
all XIS are shown (XIS0=black, XIS1=red, XIS2=green, XIS3=blue).}
\label{fig:ironline}
\vspace{0.3cm}
\end{figure}

The width, strength, and variability of the 6.4\,keV K-shell
fluorescent line of cold iron is one of the most powerful probes of
cold gas in the circumnuclear environment of an X-ray luminous AGN.
The {\it Suzaku} spectrum of \ngc\ shows the most robust evidence to
date for this iron line in this source (Fig.~\ref{fig:ironline}).
Adding a Gaussian line to the Canonical Spectral Model leads to a very
significant improvement in the goodness-of-fit ($\Delta\chi^2=-40$ for
three additional model parameters) and gives a line energy, width,
flux and equivalent width of $E=6.42\pm 0.03\keV$, $\sigma<0.07\keV$
(corresponding to a full width half maximum, FWHM$<1.1\times
10^4\kmps$), $F_{\rm K\alpha}=(6.0^{+1.9}_{-1.6})\times 10^{-6}\,{\rm
ph}\,{\rm cm}^{-2}\,{\rm s}^{-1}$ and $W_{K\alpha}=45\pm 17\eV$,
respectively.  Assuming Keplerian orbits in an edge-on accretion disk,
the limit on the FWHM corresponds to $r>3\times 10^3\,r_g$ ($6\times
10^{-3}\pc$).

We note that the equivalent width of the iron line is entirely
consistent with the limits on reflection reported in
Section~\ref{sec:broadspec}.  If, for now, we assume that the iron
line is produced by isotropic illumination of a planar optically-thick
structure, we can use the relations reported in Matt, Fabian \&
Reynolds (1997) to infer that the solid angle of the reflector
as seen by the X-ray source satisfies
\begin{equation}
\frac{\Omega}{2\pi}\approx 0.25\frac{\ln 2}{\cos\theta\,\ln(1+1/\cos\theta)},
\end{equation}
where $\theta$ is the inclination of the slab, we have assumed Anders
\& Grevesse (1989) abundances.  Note that the quantity $\Omega/2\pi$
can be compared directly with the reflection fraction quoted in
Section~\ref{sec:broadspec}.  A more sophisticated treatment of the
iron line strength expected via reflection from the surface of the
warped accretion disk in NGC~4258 (including all of the relevant
geometric effects) will be deferred to Section~4.1.

Once the narrow iron line component has been modeled, there is no
evidence in the XIS spectra for an additional broad/relativistic iron
line from the inner accretion disk.  However, due to the high
inclination angle of the inner disk the limits are not strong.
Including a Schwarzschild disk-line (modeled using the fully
relativistic code of Brenneman \& Reynolds [2006]) with a rest-frame
energy of $E=6.4\keV$ and an emissivity profile $\epsilon\sim r^{-3}$
between $r_{\rm in}=6r_g$ and $r_{\rm out}=1000r_g$, we derive an
upper limit on the equivalent width of any broad iron line of $W_{\rm
broad}<180\eV$ assuming an inner disk inclination of $i=80^\circ$.  By
contrast, even if the inner disk is in an optically-thick fluorescing
state and irradiated by an isotropic X-ray source, limb-darkening
effects would likely reduce the equivalent width of the broad iron
line to below 100\,eV (Matt et al. 1997; R00).  Thus, we cannot yet
rule out the possibility of an optically-thick X-ray irradiated inner
accretion disk.

Despite the lower signal-to-noise, the iron line is also robustly
detected in the new {\it XMM-Newton} EPIC data.  Adding a Gaussian
line leads to an improvement in the goodness-of-fit of
$\Delta\chi^2=-26$ for three additional model parameters, and gives a
line energy, width, flux and equivalent width of $E=6.41\pm 0.03\keV$,
$\sigma<0.07\keV$, $F_{\rm K\alpha}=(3.3^{+1.2}_{-1.0})\times
10^{-6}\,{\rm ph}\,{\rm cm}^{-2}\,{\rm s}^{-1}$ and $W_{K\alpha}=53\pm
19\eV$, respectively.  Comparing the {\it XMM-Newton} and {\it Suzaku}
iron line fits, we see evidence for a decrease in the flux of the iron
line.  Motivated by the desire to use line variability to locate the
fluorescing matter, this result prompts us to conduct a more
systematic analysis of the long term variability of the iron line.

\subsection{Long term variability of the iron line flux}

Given the weak nature of the iron line in NGC~4258 (which is itself a
comparatively X-ray faint AGN), there are relatively few datasets
capable of providing good constraints on the line flux.  In addition
to the deep {\it Suzaku} and {\it XMM-Newton} observations presented
here, we examine five additional datasets from the HEASARC archives; a
deep {\it ASCA} observation (15--20 May 1999; 169\,ks of good data),
and four shorter {\it XMM-Newton} observations (8-Dec-2000 [21.6\,ks];
6-May-2001 [12.9\,ks]; 17-June-2001 [13.6\,ks]; 22-May-2001
[16.5\,ks]).  We note that there is also a 15.2\,ks {\it XMM-Newton}
observation on 17-Dec-2001 that we choose to ignore due to it being
severely affected by flares in the instrumental background.  The {\it
XMM-Newton} data were processed according to the description given in
\S2.  Processing of the {\it ASCA} data follows R00 except for the use
of the latest version of the FTOOLS/HEADAS package (v6.4) and
calibration files.

\begin{figure}
\vspace{0.5cm}
\centerline{
\psfig{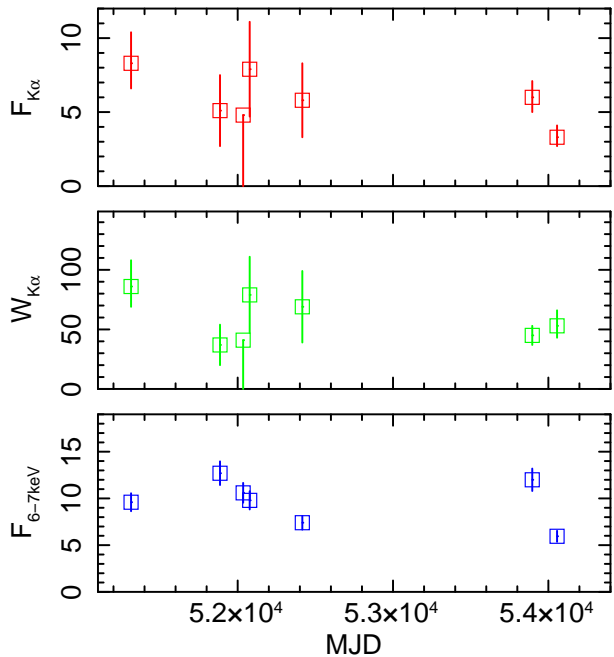}
}
\caption{Historical variability of the iron line flux (units
$10^{-6}\,{\rm ph}\,{\rm cm}^{-2}\,{\rm s}^{-1}$, equivalent width
(units $eV$), and underlying 6--7\,keV continuum (units
$10^{-13}\ergpcmsqps$).  A 10\% systematic (calibration uncertainty)
has been assumed for all continuum flux measurements.}
\label{fig:ironline_vary}
\vspace{0.3cm}
\end{figure}

Figure~\ref{fig:ironline_vary} shows the iron line flux, equivalent
width, and the 6--7\,keV continuum flux for all of the data under
consideration.  The superior statistics of the new datasets are
evident.  The hypothesis of a constant line flux can be formally
rejected at the 95\% level ($\chi^2=12.1$ for 6 degrees of freedom),
whereas the data are consistent with a constant equivalent width (a
model with $W_{\rm K\alpha}=49\eV$ gives $\chi^2=7.7$ for 6 degrees of
freedom).  The evidence for flux variability comes primarily from the
new {\it XMM-Newton} dataset (MJD~54056) which coincides with a
historical minimum in the 6-7\,keV continuum flux.

Variations of the line flux by almost a factor of two over the course
of 160 days (i.e., between the new {\it Suzaku} and {\it XMM-Newton}
observations) allows us to place interesting constraints on the
location of the fluorescing matter; the light crossing time of the
full line emitting region must be no greater than 160 days (and likely
significantly smaller).  Thus, we conclude that the line emitting
material is within a radius of $0.07\pc$ ($4\times 10^4r_g$) from the
central X-ray source.  This strongly suggests that there is cold
material (producing fluorescent iron emission) {\it within} the masing
portion of the disk (the masers appear to have an inner truncation
radius of approximately 0.13\,pc).  

\subsection{The Suzaku flare}

\begin{figure}
\vspace{0.3cm}
\centerline{
\psfig{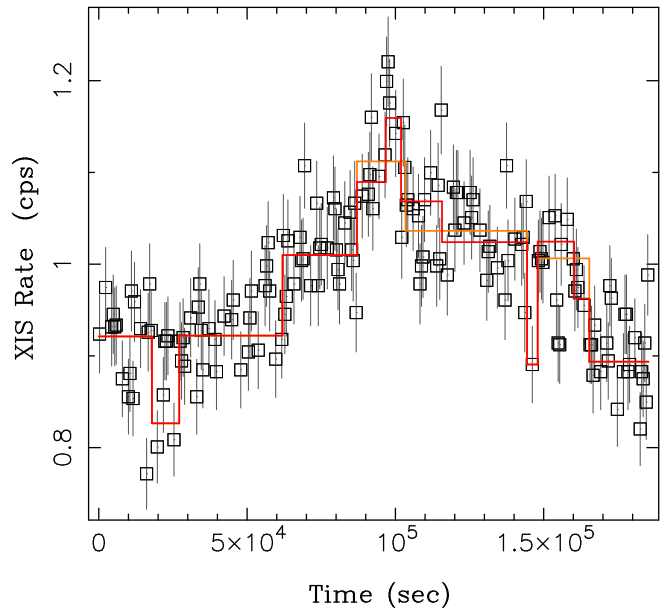}
}
\caption{Sum of the 0.3--10\,keV light curves from the Front
Illuminated XISs (XIS0,2,3).  Also shown are the results of a
``Bayesian block'' analysis with $p=3\times 10^{-5}$ (orange) and
$p=0.085$ (red).}
\vspace*{0.5cm}
\label{fig:lightcurve}
\end{figure}

Significant time variability on a range of characteristic timescales
is seen during the {\it Suzaku} observation
(Fig.~\ref{fig:lightcurve}).  A broad 100\,ksec hump is centered
within the (almost) 200\,ksec observation window.  However, much more
rapid variability is apparent.  Applying a Bayesian Blocks
analysis\footnote{http://space.mit.edu/CXC/analysis/SITAR/bb\_experiment.html}
(based upon earlier work by Scargle 1998), which searches for
flare-like structure in lightcurves dominated by Poisson statistics,
the overall lightcurve is found to be significantly variable; with
robust evidence for 10\% variations on 16\,ksec timescales (false
alarm probability $p=3\times 10^{-5}$), and good evidence for the
presence of a 5\,ksec flare ($p=8.5\times 10^{-2}$).  

Direct inspection of XIS images confirm that the variable source is
located at the nucleus of NGC~4258 to within $20^{\prime\prime}$. This
confirms and strengthens the case for rapid variability seen by {\it
BeppoSAX} (Fiore et al. 2001), and alleviates the background and
confusion concerns expressed by Fruscione et al. (2005) about the {\it
BeppoSAX} variability.  In the absence of relativistic beaming
effects, light-crossing time arguments allow one to estimate an
approximate upper limit to the size of the emitting regions; the large
amplitude 50\,ksec variability should originate from a region smaller
than about $250r_g$, whereas the corresponding limit for the rapid
5\,ksec flare is just $25r_g$.  If this variability is associated with
accretion disk processes, more stringent limits can be set by equating
these timescales with the dynamical time of a Keplerian disk
$\Omega^{-1}=\sqrt{r^3/GM}$.  Thus, under this disk-origin scenario,
the 50\,ksec variability should originate from within $r\approx 40r_g$
whereas the 5\,ksec flare should occur in the innermost disk,
$r\approx 10r_g$.

We search for any spectral changes during the flare by splitting the
{\it Suzaku} observation into a ``low-state'' and a ``high-state''.
Only the XIS data have sufficient S/N to permit this exercise.  We
define the threshold distinguishing high-state from low state to be a
total XIS count rate of $1$\,cps.  Using the Canonical Spectral Model,
the low-state XIS0--3 spectra are fit jointly with the high-state
XIS0--3 spectra.  The parameters describing the soft thermal plasma
emission are fixed between the low- and high-state spectral models.
Initially, we also assume that the photon index of the AGN emission
and the intrinsic absorption column were also the same for the two
states, with only the normalizations of the AGN power-laws being
allowed to float between the low- and high-state spectral models.  The
resulting best-fit parameters are identical to that found for the full
time-averaged spectrum.  We then allow the photon index and the
absorbing column density to be different between the low- and
high-state.  While there was a hint that both the photon index and the
absorbing column density increased in the high-state, the formal
improvement in the goodness of fit is {\it not} statistically
significant ($\Delta\chi^2=-5$ for the additional of two new model
parameters).  Thus, direct spectral modeling of the low- and
high-state spectra does not reveal robust evidence of spectral
variability.

An alternative methodology is to construct and then model the
high$-$low difference spectrum; this is a formally correct procedure
(and yields meaningful results) if the physical difference between the
high- and low-state is the addition of a new emission component.  We
find that the difference spectrum has the form of a pure absorbed
power-law; this explicitly demonstrates that, as expected, the soft
X-ray thermal plasma component has not changed between the high- and
low-state spectra.  Interestingly, both the absorption column and
photon index characterizing this variable emission differ from that
found in the time-averaged spectrum, $N_{\rm
H}=(1.58^{+0.37}_{-0.31})\times 10^{23}\,{\rm cm}^{-2}$ and
$\Gamma=2.4\pm 0.5$ [compared with $N_{\rm
H}=(9.2^{+0.4}_{-0.3})\times 10^{22}$ and
$\Gamma=1.75^{+0.05}_{-0.04}$ for the time-averaged spectrum.]  While
these results must be taken as tentative, it appears that the
difference spectra are revealing spectral changes that are too subtle
to be revealed by direct modeling of the low- and high-state spectra
alone.

\section{Discussion}

\subsection{The origin of the fluorescent iron line}

Our study has found the first significant evidence of iron line flux
variability in this source.  Combined with the limits on the velocity
width of the iron line, the variability allows us to constrain the
line emitting region to the range $3\times 10^3r_g<r<4\times 10^4r_g$
($6\times 10^{-3}\pc<r<7\times 10^{-2}\pc$).  The most obvious
candidate for the fluorescing matter on these spatial scales is the
accretion disk itself.

\begin{figure}
\centerline{
\psfig{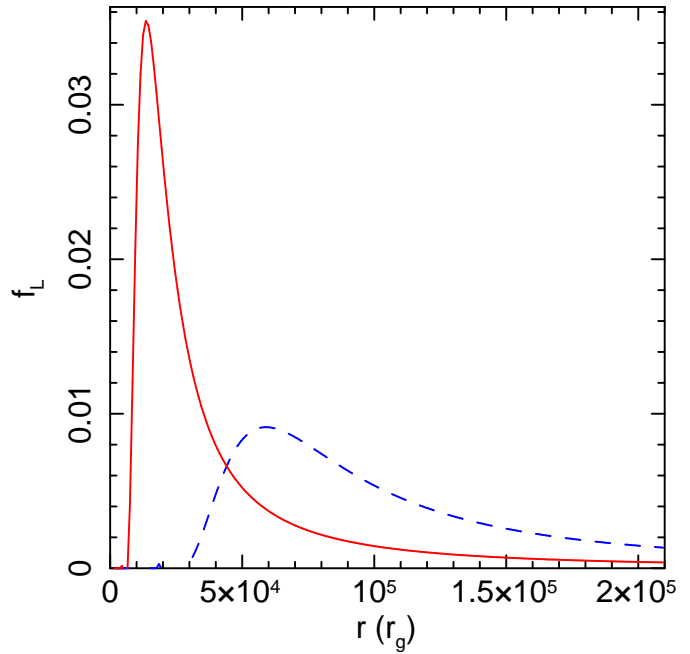}
}
\caption{Radial distribution of iron line emission in the overall best
fitting warped disk model of M08 (dashed blue line) and the best
fitting gravitationally-stable model of M08 (solid red line).  The
quantity shown here $f_L(r)$ is the total observed line flux (in
arbitrary units) from a given radius in the accretion disk.}
\label{fig:warped_results}
\end{figure}

Assuming an isotropic X-ray source at the center of the accretion
disk, the expected iron line from the disk can be estimated once we
have a model for the 3-dimensional geometry of the warped disk
relative to the observer.  Martin (2008; M08) has recently described
the maser data for NGC~4258 accretion disk using a model of a disk
warped by the General Relativistic frame-dragging effects of central
rotating black hole (Bardeen \& Petterson 1975).  As described in the
Appendix, we have calculated the expected iron line equivalent width
and the radial distribution of iron line emission for both the overall
best fitting warped disk model of M08, and the best fitting warped
disk model of M08 that is stable to self-gravity.  These two warped
disk models have a total warp angle between the inner and outer disk
of $\eta=45.8^\circ$ and $85.9^\circ$, respectively. Assuming solar
abundances (Anders \& Grevesse 1989), the $\eta=45.8^\circ$ model
gives an iron line that is both too weak ($W_{\rm K\alpha}=25\eV$) and
originates from too far out in the disk to be compatible with the line
variability ($5\times 10^4-1\times 10^5r_g$; see
Fig.~\ref{fig:warped_results}).  On the other hand, the gravitational
stable warped disk of M08 predicts an iron line of the correct
strength ($W_{\rm K\alpha}=51\eV$), originating from a region in the
disk that is compatible with both the line width and line variability
constraints ($(1-4)\times 10^4r_g$; Fig.~\ref{fig:warped_results}).
We note that in addition to the high viewing inclination, the geometry
of the warped disk implies that most of the observed iron line
emission is driven by highly oblique irradiation of the disk surface.
Thus, it is important to account for the dependence of the iron line
photon production on the angle of incidence of the irradiating X-ray
continuum (see Appendix).  The high viewing inclination and the
oblique irradiation means that simple estimates of the line
strength based on covering fraction arguments are inappropriate.

\begin{figure*}
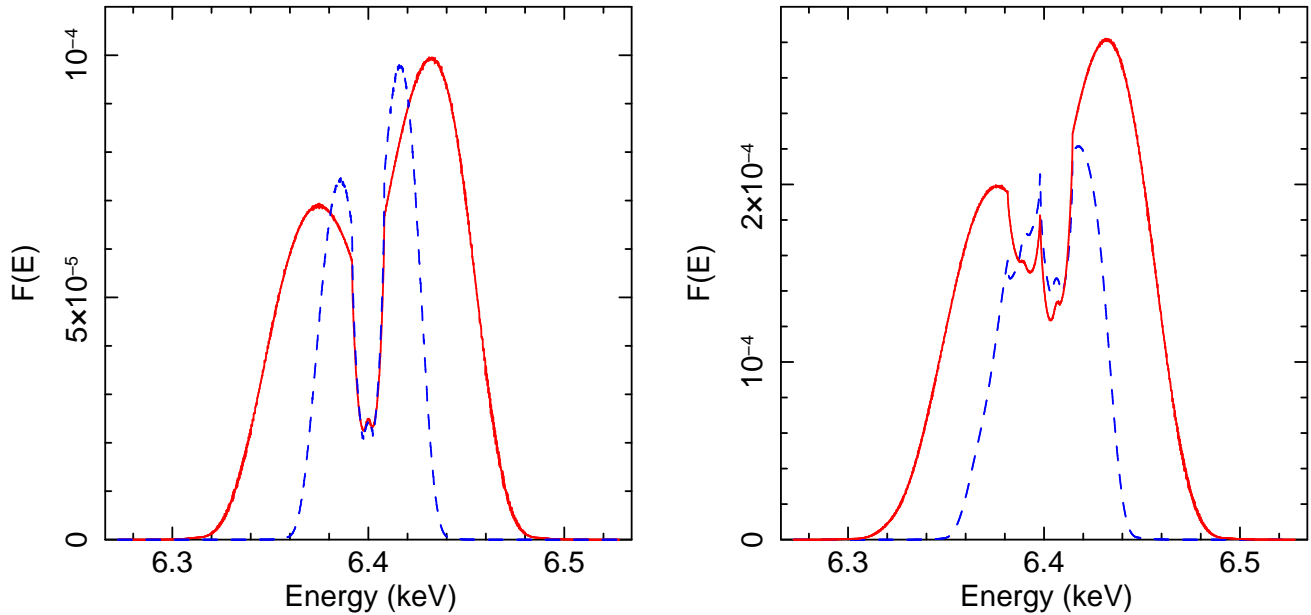

\hbox{
\psfig{figure=f6a.ps,width=0.45\textwidth,angle=270}
\hspace{0.5cm}
\psfig{figure=f6b.ps,width=0.45\textwidth,angle=270}
}
\caption{Predicted iron line profiles for the best fitting warped disk
model of M08 (dashed blue line) as well as the best fitting
gravitationally stable warped disk model (solid red line).  The {\it
left panel} shows the line profile for a single $\delta$-function
emission line at 6.4\,keV; the symmetry of the redshifts and
blueshifts is apparent.  The {\it right panel} shows the expected
profile of the Fe-K$\alpha$ doublet (with components at 6.392\,keV and
6.405\,keV and a 1:2 branching ratio.)}
\label{fig:warped_profile}
\vspace*{0.5cm}
\end{figure*}

Thus,we have shown that one can quantitatively understand the
properties of the iron line in NGC~4258 if it originates from the
surface layers of a warped disk irradiated by a central X-ray source.
We do require, however, a severe warp inside of the masing region of
the disk of the kind possessed by the M08 best-fitting gravitationally
stable model.  Interestingly, as discussed in M08, this model implies a
significant mis-alignment between the inner accretion disk and the
jet.

The warped disk hypothesis for the origin of the iron line in NGC~4258
can be tested by future X-ray spectroscopy by searching for the
predicted asymmetric double-peaked line profile.
Fig.~\ref{fig:warped_profile} shows the predicted line profiles from
the two warped disk models of M08, both for a perfect
$\delta$-function emission line at 6.4\,keV (left panel) and the real
Fe~K$\alpha$ doublet with component energies of 6.392\,keV and
6.405\,keV and a 1:2 branching ratio (right panel).  Future
high-resolution spectrographs (such as the micro-calorimeters on
{\it Astro-H} or {\it International X-ray Observatory}) will be able to
easily revolve the predicted line profiles.

\subsection{The circumnuclear environment of NGC~4258}

As discussed above, most if not all of the iron line emission in
NGC~4258 can be understood as originating from the warped accretion
disk.  This leaves very little room for any iron fluorescence from
other (non-disk) cold matter in the system such as the putative
molecular torus of unified Seyfert schemes.  Indeed, the circumnuclear
environment of this black hole does appear to be significantly
``cleaner'' than the vast majority of Seyfert-2 galaxies.  Dadina
(2008) examines a sample of 62 Seyfert 2 galaxies observed with {\it
BeppoSAX} and finds a mean reflection fraction of ${\cal R}=0.87\pm
0.14$ (assuming face-on reflection) and a mean iron line equivalent
width of $W_{K\alpha}=693\pm 195\eV$.  The strong iron line and
Compton reflection signatures in most Seyfert 2 nuclei are readily
understood as originating from the obscuring molecular torus (Krolik,
Madau \& Zycki 1994).  For NGC~4258, the fact that our almost edge-on
view of the central engine suffers absorption at ``only'' the $\sim
10^{23}\pcmsq$ level already rules out a Compton-thick torus that is
aligned with the accretion disk.  Furthermore, the very weak X-ray
reprocessing features (${\cal R}<0.43$ and $W_{K\alpha}=45\pm 17\eV$
from our time-averaged {\it Suzaku} data) rules out the presence of
even a misaligned geometrically-thick Compton-thick torus.

\subsection{A comparison with M81* and other AGN}

It is interesting to compare NGC 4258 with the LLAGN M81*, since both
seem to have very clean nuclear environments.  M81 is classified as a
Seyfert 1.8 / Low-Ionization Emission Line Region (LINER) galaxy, with
a central black hole mass of $M = 7 \times 10^7 M_\odot$ (Devereux et
al. 2003) accreting at $L \simeq 10^{-5} L_{\rm Edd}$.  M81* shows
evidence of a one-sided radio jet emanating from the nucleus
(Bietenholz, Bartel \& Rupen 2000).  High-resolution X-ray
spectroscopy of M81* with \emph{Chandra} (Young et al. 2007) reveals a
number of emission lines, including Fe K$\alpha$, and velocity
broadened Si K$\alpha$ and other thermal lines.  The Fe K$\alpha$ line
in M81* has an EW of $47^{+25}_{-24}$ eV, almost identical to that of
NGC 4258, and there is no evidence of a broad iron line.  Furthermore,
the broadened Si K$\alpha$ fluorescence line in M81* is consistent
with originating in a disk at $r \sim 10^4 r_g$, assuming the
inclination angle is the same as the disk observed with HST, $i =
14^\circ$ (Devereux et al. 2003).  The broadened thermal lines in M81*
are consistent with originating at $r < 10^{4-5} r_g$, suggesting that
there is hot thermal gas at small radii, possibly in the form of a
radiatively inefficient accretion flow or the base of a jet (Markoff
et al. 2008).

The inclination angles of the outer accretion disk in NGC 4258 ($ \sim
80^\circ$), and in M81, ($14^\circ$) are significantly different, and
this may account for the fact that NGC 4258 has a much larger column
density ($\sim 10^{23}$ cm$^{-2}$) than M81 ($\sim 10^{21}$
cm$^{-2}$), that maser emission is not seen in M81, and the difference
in the obscuration of the broad line region (i.e., the Seyfert types).
In M81*, we do not know the geometry of the disk well enough to
calculate the strength of the Fe K$\alpha$ line that it would produce.
Apart from their different inclination angles, the accretion flows in
NGC 4258 and M81* seem to be remarkably similar.

Although we have noted the similarities between the circumnuclear
environments of NGC4258 and M81*, it is interesting that these results
on NGC4258 and M81* appear to run counter to some general behaviour of
the AGN population.  In particular, it has been noted that the
strength of the iron emission line across the AGN population is
anticorrelated with both the X-ray luminosity (Iwasawa \& Taniguchi
1993) and the Eddington ratio (Winter et al. 2008).  Using the
Swift-BAT AGN survey, Winter et al. (2008) show that LLAGN with X-ray
Eddington ratios comparable to NGC~4258 ($L_X/L_{\rm Edd}\sim
10^{-5}$) typically possess iron line equivalent widths in excess of
500\,eV.  Thus, on the basis of the iron line equivalent width, the
circumnuclear environments of NGC~4258 and M81* appear to have
significantly less cold gas than the average LLAGN.

\section{Conclusions}

Using {\it Suzaku}, {\it XMM-Newton} and {\it Swift}, we have obtained
an unprecedented view of the active nucleus in NGC~4258.  Our
principal results are:
\begin{enumerate}
\item Comparing the {\it Suzaku} data with {\it XMM-Newton} data taken
160 days later, we detect robust flux variability of the 6.4\,keV iron
line for the first time.  Together with constraints on the velocity
width of the line, and assuming Keplerian motion about the central
black hole, we can place the iron line emitting region to be between
$3\times 10^3r_g$ and $4\times 10^4r_g$.
\item We show that the strength, velocity width and time variability
of the iron line can be explained by a model in which the line
originates from the surface of a warped accretion disk.  In
particular, we present explicit calculations of the expected iron line
from a disk warped by Lens-Thirring precession from a severely
misaligned central black hole.
\item During our {\it Suzaku} observation, we detect high amplitude
intraday variability, with fluctuations on timescales as short as
5\,ks.  Corresponding light travel time arguments suggest that the
emission region is smaller than $25r_g$.  If we make the stronger
assertion that this timescale be longer than the dynamical timescale
of the accretion disk at the location it is produced, the upper limit
on the radius of the emission is $10r_g$.
\item In stark contrast with the vast majority of other Seyfert 2
galaxies, there are no indications of a Compton-thick obscuring torus;
the weak iron line and the lack of reflection all point to a
circumnuclear environment that is remarkable clean of cold gas.  As
pointed out by Herrnstein et al. (2005), the intrinsic absorption that
we do see in the X-ray spectrum may well arise in the outer layers of
the warped geometrically-thin accretion disk, further reducing the
need for any cold structure other than the accretion disk itself.
\item We highlight the similarities in the circumnuclear environments
of NGC~4258 and another LLAGN, M81*.  However, we also note that the
remarkably clean circumnuclear environment found in these two LLAGN
stand in contrast to the vast majority of LLAGN.
\end{enumerate}

We thank Richard Mushotzky for stimulating conversations throughout
this work.  CSR thanks the NASA {\it Suzaku} and {\it XMM-Newton}
Guest Observer Programs for support under grants NNX06A135G and
NNX07AE97G.

\section*{References}

{\small

\noindent Anders E., Grevesse N., 1989, Geochimica et Cosmochimica
Acta, 53, 197

\noindent Bardeen J.M., Petterson J.A., 1975, ApJL, 195, L65

\noindent Brenneman L.W., Reynolds C.S., 2006, ApJ, 652, 1028

\noindent Cecil G., Wilson A.~S., De Pree C., 1995, ApJ, 440, 181

\noindent Chiang J. et al., 2000, ApJ, 528, 292

\noindent Dadina M., 2007, A\&A, 461, 1209

\noindent Devereux N., Ford H., Tsvetanov Z., Jacoby G., 2003, AJ,
125, 1226

\noindent Dickey J.M., Lockman F.J., 1990, ARA\&A, 28, 215

\noindent Fabbiano G., Kim D., Trincheieri G., 1992, ApJS, 80, 531

\noindent Fiore F., et al., 2001, ApJ, 556, 150

\noindent Fruscione A., Greenhill L.J., Filippenko A.V., Moran J.M.,
Hernnstein J.R., Galle, E., 2005, ApJ, 624, 103

\noindent Gammie C.F., Narayan R., Blandford R.D., 1999, ApJ, 516, 177

\noindent George I.M., Fabian A.C., 1991, MNRAS, 249, 352

\noindent Herrnstein J.R., et al., 1999, Nature, 400, 539

\noindent Herrnstein J.R., Moran J.M., Greenhill L.J., Trotter A.S.,
2005, ApJ, 629, 719

\noindent Kaastra J.S., 1992, An X-ray Spectral Code for Optically
Thin Plasmas (Internal SRON-Leiden Report, updated version 2.0)

\noindent Krolik J.H., Madau P., Zycki P.T., 1994, ApJL, 420, L57

\noindent Iwasawa K., Taniguchi Y., 1993, ApJ, 1993, 413, L15

\noindent Lasota J.P., Abramowicz M.A., Chen X., Krolik J., Narayan
R., Yi L., 1996, ApJ, 462, 142

\noindent Lee J.C., Fabian A.C., Reynolds C.S., Brandt W.N., Iwasawa
K., 2000, MNRAS, 318, 857

\noindent Liedahl D.A., Osterheld A.L., Goldstein W.H., 1995, ApJ, 438, L115

\noindent Makishima K. et al., 1994, PASJ, 46, L77

\noindent Magdziarz P., Zdziarski A.A., 1995, MNRAS, 273, 837

\noindent Markoff S. et al., 2008, ApJ, in press

\noindent Martin R.G., 2008, MNRAS, in press (arXiv0804.1013) [M08]

\noindent Martin R.G., Pringle J.E., Tout C.A., 2007, MNRAS, 381, 1617

\noindent Matt G., Fabian A.C., Reynolds C.S., 1997, MNRAS, 289, 175

\noindent Mewe R., Gronenschild E.H.B.M., van den Oord G.H.J., 1985,
A\&AS, 62, 197

\noindent Miyoshi M., Moran J., Herrnstein J., Greenhill L., Nakai N.,
iamond P., Inoue M., 1995, Nature, 373, 127

\noindent Neufeld D., Maloney P.R., 1995, ApJ, 447, L17

\noindent Pietsch W., Vogler A., Kahabka P., Klein U., 1994, A\&A,
284, 386

\noindent Pietsch W., Read A.M., 2002, A\&A, 384, 793

\noindent Reynolds C.S., Nowak M.A., Maloney P.R., 2000, ApJ, 540, 143 [R00]

\noindent Reynolds C.S., Nowak M.A., 2003, Phys. Rep., 377, 389

\noindent Scargle J.D., 1998, ApJ, 504, 405

\noindent Scheuer P.A.G., Feiler R., 1996, MNRAS, 282, 291

\noindent Tueller J., et al., ApJ, in press (arXiv:0711.4130)

\noindent Vogler A., Pietsch W., 1999, A\&A, 352, 64

\noindent Wilson A.S., Yang Y., Cecil G., 2001, ApJ, 560, 689

\noindent Winter L.M., Mushotzky R.F., Reynolds C.S., Tueller J.,
2008, ApJ, in press

\noindent Yuan F., Markoff S., Falcke H., Biermann P.L., 2002, A\&A,
391, 139

\noindent Young A.J., Wilson A.S., 2004, ApJ, 601, 133

\noindent Young A.J., Nowak M.A., Markoff S., Marshall H.L., Canizares
C.R., 2007, ApJ, 669, 830 }

\appendix

\section{Iron line emission from a centrally illuminated warped disk}

In this Appendix, we describe our calculation of the fluorescent iron
line profiles from a centrally illuminated warped accretion disk.  We
base our considerations on the warped disk model for NGC~4258 of
Martin (2008; also see Martin, Pringle \& Tout 2007).  Martin et
al. (2008; hereafter M08) use the formulism of Scheuer \& Feiler
(1996) to compute the geometry of a steady-state, geometically-thin
viscous disk that is warped due to a misalignment between the angular
momentum of the large scale accretion flow and the spin axis of the
central black hole.  The interesting parameters of the M08 disk model
are the misalignment angle between the angular momentum of the large
scale accretion flow and the spin axis of the central black hole,
$\eta$, the warp radius $R_{\rm warp}$, and the power-law index
describing the radial dependence of the surface density $b$ (with
surface density following $\Sigma\propto R^{-b}$.  Setting $b=2$, M08
show that such a model describes the geometry of the high-velocity
masers in NGC~4258.  Although some degeneracy exists, M08 described
two ``best fitting'' models; the overall best-fit has
$\eta=45.8^\circ$ and $R_{\rm warp}=1.6\times 10^5r_g$ (corresponding
to 5.8\,mas), whereas the best-fitting model given the constrain of a
gravitationally-stable disk is $\eta=85.9^\circ$ and $R_{\rm
warp}=2.47\times 10^4r_g$ (corresponding to 1.3\,mas).

A basic premise of this approach is that the disk at a given radius is
a flat ring centered on the black hole.  The different annuli exchange
angular momentum via a process modeled as shear viscosity.  Let ${\bf
l}(R)$ be a unit vector pointing in the direction of the angular
momentum of the disk flow at radius $R$.  In a Cartesian coordinate
system in which the black hole spin is aligned with the $z$-axis, the
fact that the innermost portions of the disk are aligned with the
black hole spin (Bardeen \& Petterson 1975) imply that ${\bf
l}(R)\rightarrow (0,0,1)$ as $r\rightarrow 0$.  We also have ${\bf
l}(R)\rightarrow (\sin\eta,0,\cos\eta)$ as $r\rightarrow \infty$,
where we have aligned our coordinate system such that the angular
momentum of the large scale flow lies in the $x-z$ plane.  As
discussed in Martin, Pringle \& Tout (2007) and M08, the disk warp is
then described by
\begin{equation}
W(R)=\frac{2W_\infty}{\Gamma(\frac{1}{6})}\frac{(-i)^{1/12}}{3^{1/6}}\left(\frac{R_{\rm warp}}{R}\right)^{1/4}\,K_{1/6}\left[\frac{\sqrt{2}}{3}(1-i)\left(\frac{R}{R_{\rm warp}}\right)^{-3/2}\right],
\end{equation}
where $W=l_x+il_y$ is a complex representation of the unit vector
${\bf l}$, $W_\infty=\sin\eta$, $K_{1/6}$ is the modified Bessel
function of order $1/6$, and $\Gamma$ is the gamma-function.   

Within the context of their models, M08 find good fits to the maser
data assuming that both the black hole spin axis and the angular
momentum of the large scale accretion flow lie in the plane of the
sky.  Hence, in the Cartesian coordinate system defined above, our
line of sight to the disk is directed along the y-axis, i.e., in the
direction of the unit vector ${\bf e}_y$.

Given the disk geometry relative to the observer, we now proceed to
compute the X-ray irradiation of the disk surface and the resulting
iron line emission.  Suppose that the primary X-ray source is located
at the center of the accretion disk, and emits continuum photons
isotropically at a rate $N_c(E)dE=N_0(E/E_0)^{-\Gamma}$\,dE where, from
now on, $\Gamma$ shall refer to the photon index of the power-law
source.  Following Pringle (1996), we define the angles $\gamma(R)$
and $\beta(R)$ such that 
\begin{equation}
{\bf l}=(\cos\gamma\sin\beta,
\sin\gamma\sin\beta, \cos\beta),
\end{equation}
i.e., $\beta$ is the local angle of tilt of the disk relative to the
$z$-axis and $\gamma$ marks the position of the decending node of the
disk.  We shall also establish a polar coordinate system $(R,\phi)$ on
the surface of the warped disk.  We shall denote as ${\bf x}$ the
position vector of the point on the disk with coordinates $(R,\phi)$;
the corresponding unit vector shall be denoted $\hat{\bf x}$. Pringle
(1996) gives
\begin{equation}
{\bf x}=R(\cos\phi\sin\gamma+\sin\phi\cos\gamma\cos\beta,
\sin\phi\sin\gamma\cos\beta-\cos\phi\cos\gamma, -\sin\phi\sin\beta)
\end{equation}

From eqn.~(2.13) of Pringle (1996), the rate at which a patch ${\bf
dS}$ of the disk surface intercepts photons from the central primary
source (neglecting possible shadowing) is
\begin{equation}
dN=\frac{N_c\,dE}{4\pi R^2}|\hat{\bf x}\cdot{\bf dS}|=\frac{N_c\,dE}{4\pi R^2}|R\gamma~^\prime\cos\phi\sin\beta-R\beta^\prime\sin\phi|R\,dR\,d\phi,
\end{equation}
where $\gamma~^\prime\equiv d\gamma/dR$ and $\beta^\prime\equiv
d\beta/dR$.  The angle $\theta_0$ between the irradiating flux and
local disk normal is given by
\begin{equation}
\cos\theta_0=\frac{\hat{\bf x}\cdot {\bf dS}}{|{\bf dS}|}=\frac{R(\cos\phi\sin\beta\gamma~^\prime-\sin\phi\beta^\prime)}{[1+R^2(\cos\phi\sin\beta\gamma~^\prime-\sin\phi\beta^\prime)^2]^{1/2}}.
\end{equation}

Given the irradiating continuum photon flux and the incident angle, we
use eqns.~4-7 of George \& Fabian (1991) to compute the number of
fluorescent iron line photons generated by the patch,
$N_lR\,dR\,d\phi$.  We then apply a scaling factor of 1.3 (Matt et
al. 1997) to correct the results of George \& Fabian (1991) to those
expected for Anders \& Grevesse (1989) solar abundances.  Assuming
that these photons are emitting isotropically from the optically-thick
surface of the disk, the iron line photon emission rate from the patch
per unit solid angle towards the observer is
$\cos\theta\,N_lR\,dR\,d\phi/\pi$, where $\theta$ is the angle between
the local disk normal and the observer's line of sight given by.
\begin{equation}
\cos\theta=\frac{{\bf e}_y\cdot {\bf dS}}{|{\bf dS}|}
=\frac{l_y-R(\sin\phi\sin\gamma\cos\beta-\cos\phi\cos\gamma)(\gamma~^\prime\cos\phi\sin\beta-\beta^\prime\sin\phi)}{[1+R^2(\cos\phi\sin\beta\gamma~^\prime-\sin\phi\beta^\prime)^2]^{1/2}}.
\end{equation}

The contribution of a given radial range $r\rightarrow r+dr$ to the
observed line emission is given by
\begin{equation}
{\cal N}(R)=\int_0^{2\pi}\frac{N_l\cos\theta}{\pi}Rd\phi,
\end{equation}
and the total number of observed iron line photons is ${\cal N}~_{\rm
tot}=\int_0^{\infty}{\cal N}(R)\,dR$.  The observed equivalent width
is derived by ratioing this against the number of continuum photons
emitted per unit solid angle at the line energy,
\begin{equation}
W_{\rm K\alpha}\frac{{\cal N}~_{\rm tot}}{N_c(E_{\rm K\alpha})/4\pi}.
\end{equation}

Finally, to compute a line profile, we need to know the line-of-sight
velocity of a given patch of the disk.  Assuming Keplerian flow, the
velocity of the disk is given by 
\begin{equation}
{\bf v}={\bf l}\times \hat{{\bf x}}\sqrt\frac{GM}{R},
\end{equation}
and the observed line of sight velocity is 
\begin{equation}
v_{\rm los}={\bf e}_y\cdot
{\bf v}
=\sqrt\frac{GM}{R}[\sin\phi\cos\gamma\sin^2\beta+\cos\beta(\cos\phi\sin\gamma+\sin\phi\cos\gamma\cos\beta)].
\end{equation}

\end{document}